\documentclass[10pt,a4paper,twoside]{article}
\usepackage{epsfig}
\usepackage{baltlat6}
\usepackage{wrapfig}
\usepackage{dcolumn}
\begin{document}
\ \
\vspace{-0.5mm}

\setcounter{page}{235}
\vspace{-2mm}

\titlehead{Baltic Astronomy, vol.\ts 14, 235--244, 2005.}

\titleb{RADIATIVE TRANSFER PROBLEM IN DUSTY GALAXIES:\\
EFFECTS OF NON-ISOTROPIC MULTIPLE SCATTERING}

\begin{authorl}
\authorb{D. Semionov}{} and
\authorb{V. Vansevi{\v c}ius}{}
\end{authorl}

\moveright-3.2mm
\vbox{
\begin{addressl}
\addressb{}{Institute of Physics,
Savanori{\c u} 231, Vilnius LT-02300, Lithuania}
\end{addressl}
}

\submitb{Received 2005 April 5; revised 2005 May 24}

\begin{summary}
We investigate the effects of multiple anisotropic light
scattering by interstellar dust particles on photometric
profiles of disk galaxy models computed using an iterative
ray-tracing radiative transfer modeling code. It is shown
that anisotropic scattering must be fully accounted for in
at least the first scattering event for all considered cases.
At the same time scattering terms of the order higher than 5th
can be safely approximated using the isotropic phase function.
The effects of isotropic approximation are most significant
in the galaxy models seen face-on with dust population extending
beyond the stellar disk, therefore the applicability of
isotropic approximation to a higher-order (2 to 5) scattering
terms depends on the assumed galaxy model parameters,
namely, the optical depth, scattering albedo and the
distribution of interstellar dust with respect to stars.
\end{summary}


\resthead{Effects of non-isotropic multiple scattering}{D. Semionov, V. Vansevi{\v c}ius}

\vskip1mm
\sectionb{1}{INTRODUCTION}
\vskip-1mm

Optical interstellar dust grain properties still remain an
uncertain factor in our understanding of star formation
history in galaxies. Existing models and theories of interstellar
dust are primarily intended to describe the optical properties
of grains by means of size, composition and relative abundance
of different grain types, with limited connection to the general
chemical evolution of interstellar matter (ISM) (Li \& Greenberg
2003). At the same time, observations of the Milky Way and
external galaxies reveal a multitude of existing extinction laws and
relative distributions of stellar populations and ISM. A
straightforward attempt to reproduce full complexity of the
processes in ISM in order to model the observed galactic spectral
energy distribution (SED) introduces a large number of free
parameters, resulting in a significantly degenerate solution
(e.g., Silva \etal\ 1998). Considering the difficulty of practical
and theoretical determination of the optical properties and the
lack of the unified theory of ISM, an attempt to analyze a
simplified model to discern relative influence of stellar and
dust content and their distribution might prove to be fruitful
(e.g., Ferrara \etal\ 1999).

A canonical formulation of the radiative transfer (RT) problem
in the form (Chandrasekhar 1960)

\begin{equation}
{dI_\lambda \over ds} = - \kappa_\lambda I_\lambda + j_\lambda +
\kappa_\lambda {\omega_\lambda \over 4 \pi}
\int I_\lambda \Phi_\lambda(\Omega) d \Omega,
\end{equation}

\noindent defines three main optical properties of intervening
matter: extinction mass coefficient $\kappa_\lambda$, scattering
albedo $\omega_\lambda$ and scattering phase function $\Phi$,
commonly described by means of phase function parameter $g_\lambda$
(Henyey \& Greenstein 1941). While the values of the first
two parameters might be obtained from the total emissivity, the
absorbed (and re-radiated) energy and the observed SED, the
determination of phase function anisotropy is less certain and
usually depends on assumptions about structure, chemical
composition and size distribution of interstellar dust grains.

In this paper we attempt to estimate the influence of scattering
phase function anisotropy on the SEDs of model disk galaxies. The
description of the RT code, used to compute model SEDs, and the
assumed galaxy model parameters are presented in Section 2. The
results, obtained using the isotropic approximation are compared
with those from fully anisotropic calculations in Section 3. In
the same section we discuss isotropic approximation applicability
limits.

\vskip1mm
\sectionb{2}{MODELS AND METHODS}
\vskip-1mm

To test the influence of the anisotropic scattering phase function,
the RT problem has been solved for the six pure disk galaxy model
families S1--S6 with varying relative star and dust distributions.
Both stars and dust are assumed to have double-exponential
distribution of luminosty and mass, respectively:
\vskip-2mm

\begin{equation}
\rho(r,z) = \rho_0 \exp \left( -{r\over r_{\rm eff}} -{\left| z \right| \over z_{\rm eff}}\right),
\end{equation}

\noindent $r_{\rm eff}$ and $z_{\rm eff}$ being the effective
scale-length and scale-height for the appropriate distribution. The
common
effective scale-length of stellar and dust distribution was assumed
to be constant, while the effective scale-height of dust distribution
$z^d_{\rm eff}$ was varied with respect to the effective scale-height
of stellar disk to represent three commonly used cases -- ``dust
within stellar disk'' ($z^d_{\rm eff} = 0.5 z_{\rm eff}$, models S1
and S4), ``well-mixed dust and stars'' ($z^d_{\rm eff} = z_{\rm eff}$,
models S2 and S5) and ``dust enveloping stellar disk''
($z^d_{\rm eff} = 2 z_{\rm eff}$, models S3 and S6). We have used two
values of total dust mass, corresponding to the optical depth to the
model center in $V$ passband, measured perpendicularly to the disk
plane: $\tau_V = 1$ for models S1--S3 and 10 for models S4--S6.

\begin{wrapfigure}{l}[0pt]{6.0cm}
\vbox{\footnotesize
\begin{tabular}{lrrrrrr}
\multicolumn{7}{c}{\parbox{5.5cm}{\baselineskip=8pt
~~{\smallbf Table 1.}{\small\ Galaxy model parameters.}}}\\
\tablerule
  & S1 & S2 & S3 & S4 & S5 & S6 \\
\tablerule
$z^d_{\rm eff}/z_{\rm eff}$ & 0.5 & 1 & 2 & 0.5 & 1  & 2  \\
$\tau_V$                    & 1   & 1 & 1 & 10  & 10 & 10 \\
\tablerule
\end{tabular}
}
\end{wrapfigure}

Each model family in S1--S6 contains a sequence of five models computed
with eight scattering iterations and denoted M0:8, M1:7, M3:5, M5:3 and
M8:0, with first and second digits in the model name corresponding to
the number of initial anisotropic and subsequent isotropic iterations
respectively (e.g., M0:8 being computed with fully isotropic scattering,
M1:7 has first iteration computed with anisotropic scattering and
following seven iterations using isotropic approximation, etc.).

Optical dust grain properties, shown in Figure 1, were computed using
the Laor \& Draine (1993) model, approximating the Milky Way galaxy type
extinction using a mixture of graphite and silicate grains in the
proportion of 0.47 to 0.53, with grain size distribution following the
power law $a^{-3.5}$ with lower and upper cut-off radii $a_{\rm
min}=0.005$ $\mu$m and $a_{\rm max}=0.25$ $\mu$m.  SEDs of the models
were computed using the Galactic Fog Engine RT code (GFE, Semionov \&
Vansevi{\v c}ius 2002; Semionov 2003; Semionov \& Vansevi{\v c}ius
2005), implementing an iterative 2-D ray-tracing algorithm in
axisymmetrical geometry.  The galaxy model is represented as a cylinder,
divided into 20 layers consisting of 25 concentric rings, truncated at
the radius $r_{\rm max} = 6 r_{\rm eff}$ and the height above and below
the disk central plane $z_{\rm max} = \pm 6 z_{\rm eff}$.  The GFE uses
a static ray-casting scheme to produce a sufficient sampling of the
galaxy model by a set of rays, and then solves 1-D RT problem (Eq. 1)
along each ray.  During the first iteration the initial global radiation
field of the system, $I_0$, determined from the stellar luminosity
distribution, contributes to the escaped energy (the observed SED), the
absorbed energy $E^{\rm abs}_\lambda$ and the global radiation field of
once-scattered light, $I_1$.  Higher-order ($i$th) scattering events are
accounted for by substituting $I_{i-1}$ as input to calculate $I_i$ and
accumulating obtained escaped and absorbed energy in subsequent
iterations.

\begin{wrapfigure}{i}[0pt]{65mm}
\centerline{\psfig{figure=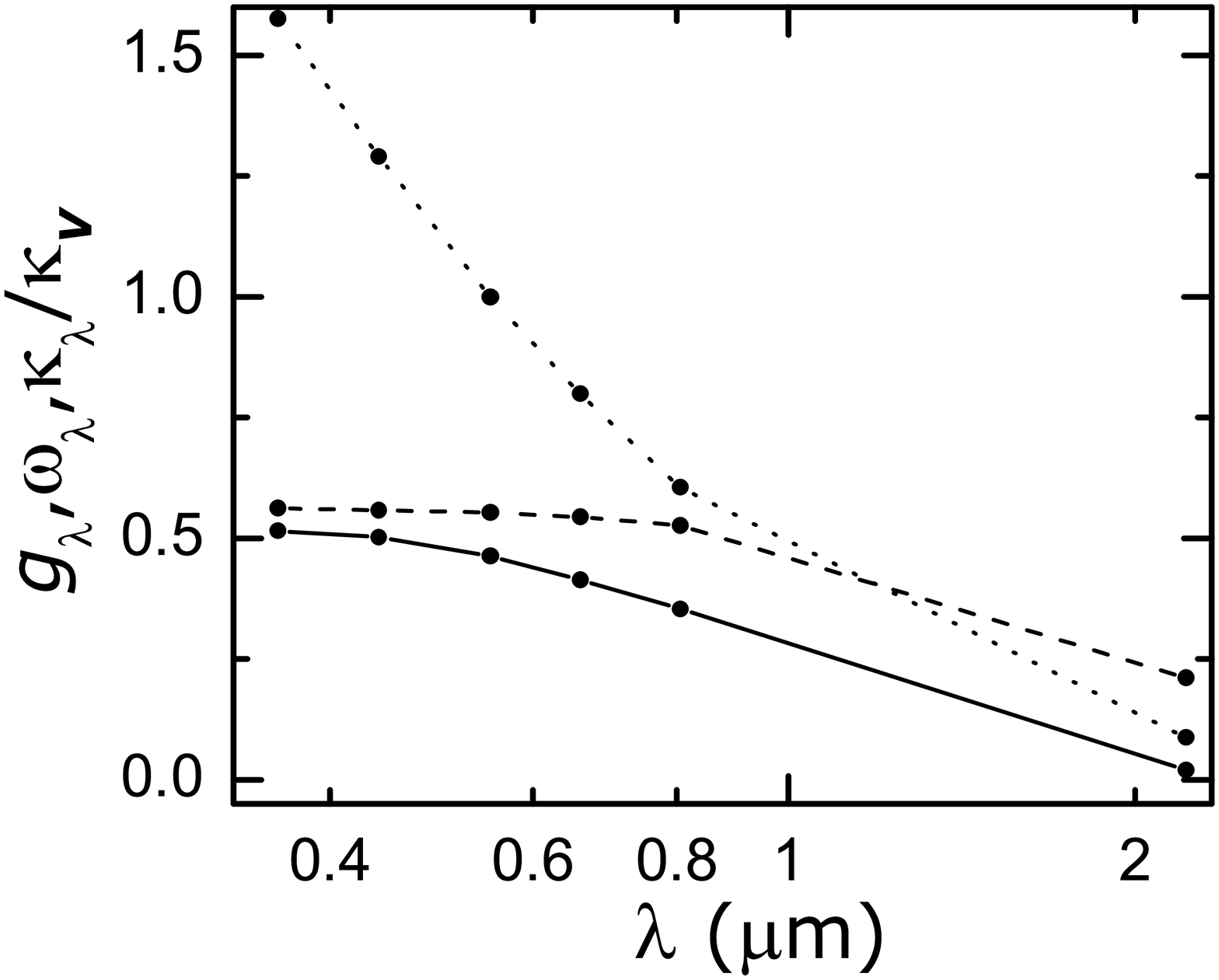,width=65truemm,angle=0,clip=}}
\captionb{1}{Assumed optical properties of interstellar grains.
Solid, dashed and dotted lines show the scattering phase function
parameter $g_\lambda$, albedo $\omega_\lambda$ and relative mass
extinction coefficient $\kappa_\lambda / \kappa_{\rm V}$,
respectively. Dots show positions of the $U$, $B$, $V$, $R$, $I$
and $K$ passbands.}
\end{wrapfigure}
\vskip1mm

For all models at all considered wavelengths the amount of total
radiative energy, remaining to be scattered after the 8th iteration, and
a total energy defect due to numerical inaccuracies were found to be
below 0.5\% and 1\% of total emitted energy, $E^{\rm tot}_\lambda$,
respectively.  The resulting attenuation curves (the ratio of the flux
``observed'' from the dusty model galaxy to the flux from the dust-free
model galaxy), $A(\lambda) = F_\tau (\lambda) / F_0 (\lambda)$, and the
spectral distribution of the normalized absorbed energy (the ratio of
the absorbed energy to the total emitted energy), $E^{\rm abs}_\lambda /
E^{\rm tot}_\lambda$, are shown in Figure 2. We have found, that the
total amount of the absorbed energy in model galaxies $E^{\rm
abs}_\lambda$ only slightly depends on $g_\lambda$, varying within 2\%
of $E^{\rm tot}_\lambda$ for all considered cases and thus cannot
indicate the effect of anisotropic scattering.

\begin{figure}
\centerline{\psfig{figure=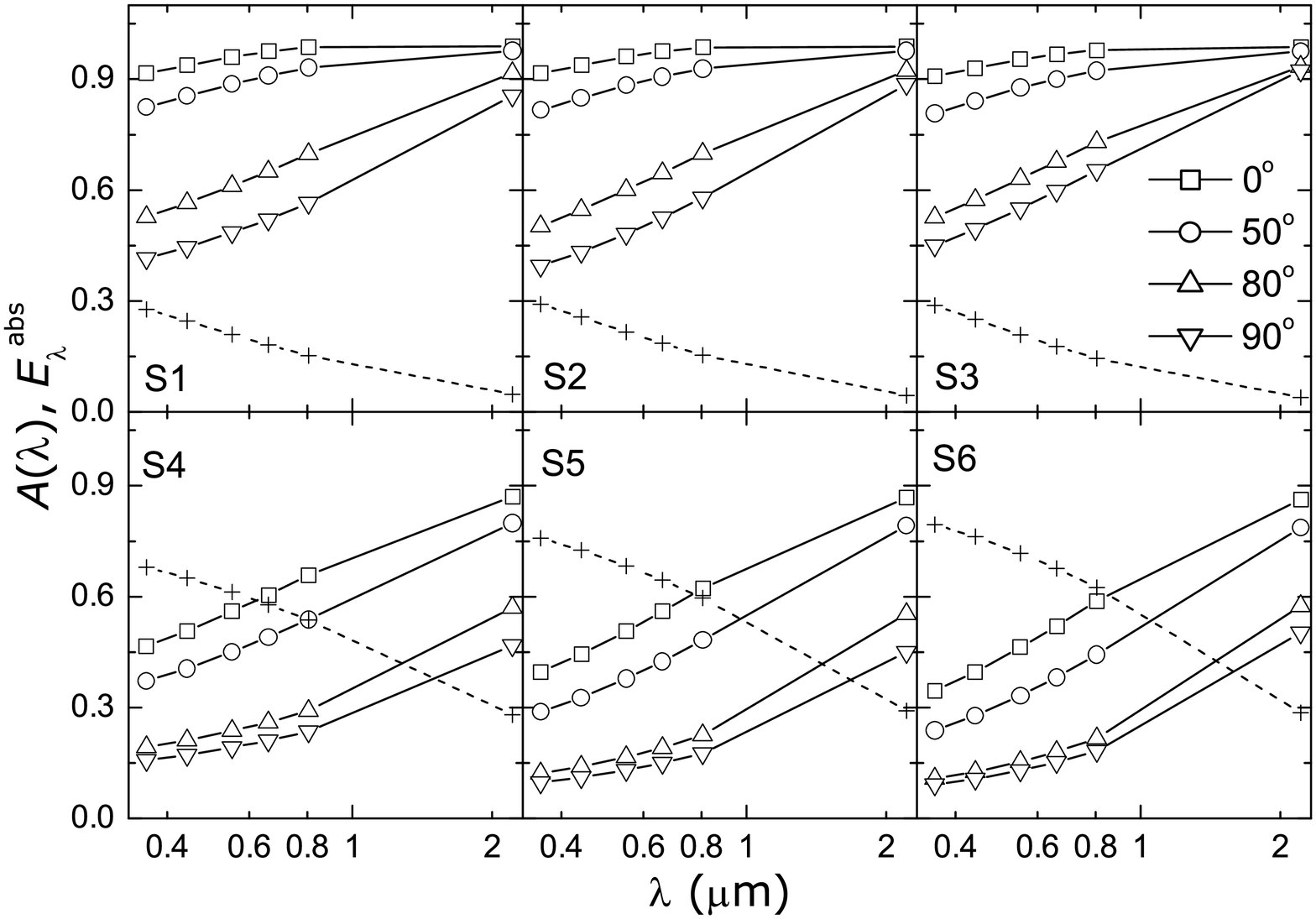,width=100mm,angle=0,clip=}}
\captionb{2}{Attenuation curves and absorbed energy for the models
S1--S6. Solid lines show attenuation curves for model galaxies with
inclinations of 0\degr, 50\degr, 80\degr\ and 90\degr, the dotted line
shows the normalized absorbed energy.}
\end{figure}

\begin{figure}
\centerline{\psfig{figure=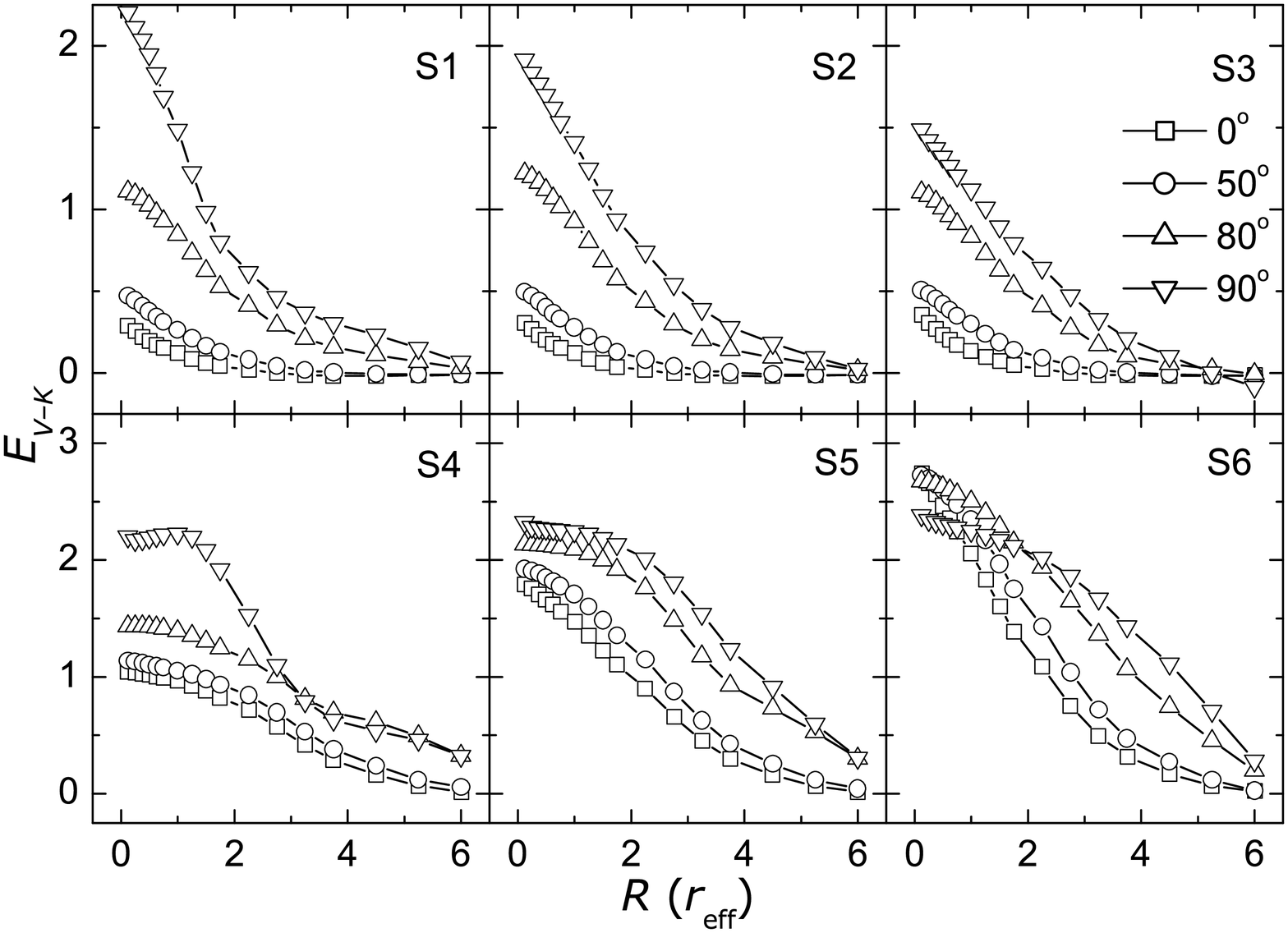,width=100mm,angle=0,clip=}}
\captionb{3}{Photometric profile of color excess
$E_{V-K}$  for the model groups S1--S6.
Lines correspond to the model galaxy inclinations of 0\degr,
50\degr, 80\degr \ and 90\degr. The aperture semi-major axis
radius is given in units of the effective stellar disk scale-length
$r_{\rm eff}$.}
\end{figure}

For all model families we have computed images with a scale of $0.1
z_{\rm eff}$ per pixel at the wavelengths corresponding to the $U$, $B$,
$V$, $R$, $I$ and $K$ passbands for model galaxy inclinations of 0\degr,
50\degr, 80\degr\ and 90\degr.  To investigate the differences in the
observed quantities of galaxy models we performed surface photometry of
these images using apertures with ellipticity determined from the model
inclination, the aperture being centered on the geometric center of the
model galaxy image, obtaining differential azimuthally-averaged color
excess profiles.  The example of the color excess $E_{V-K}$ photometric
profiles for the galaxy models M8:0, computed with fully anisotropic
scattering, is shown in Figure 3 as a function of the aperture
semi-major axis radius given in units of $r_{\rm eff}$.  Since the
observed minor axis cross-sections of galaxies are more frequently
available than multiaperture photometry or complete photometric profiles
and are widely used in determination of the effective scale-lengths and
scale-heights of stellar populations and extinction distribution, we
have extracted from the images color excess profiles along their minor
axes.  Hereafter we will use differences of photometric profiles and
minor axis cross-sections (denoted by a ``$C$'' superscript) between the
fully anisotropic models M8:0 on one side and the models computed using
varying levels of isotropic approximation:  M0:8, M1:7, M3:5 and M5:3,
on other side. For example, in the case of $E_{V-K}$ we have

\begin{equation}
\Delta E_{V-K}^{Mi} = E_{V-K}^{M\rm{8:0}} - E_{V-K}^{Mi}.
\end{equation}

\vskip-1mm

\sectionb{3}{RESULTS AND DISCUSSION}

\vskip-1mm

The results obtained allow us to distinguish two model groups according
to their effective opacity: the low opacity models S1--S4 and the high
opacity models S5 and S6. Models in the group S1--S4 display similar
isotropic approximation-induced error distributions and their absolute
values for most of the performed tests, while models S5 and S6 appear
significantly different and have to be considered separately.

\begin{figure}
\centerline{\psfig{figure=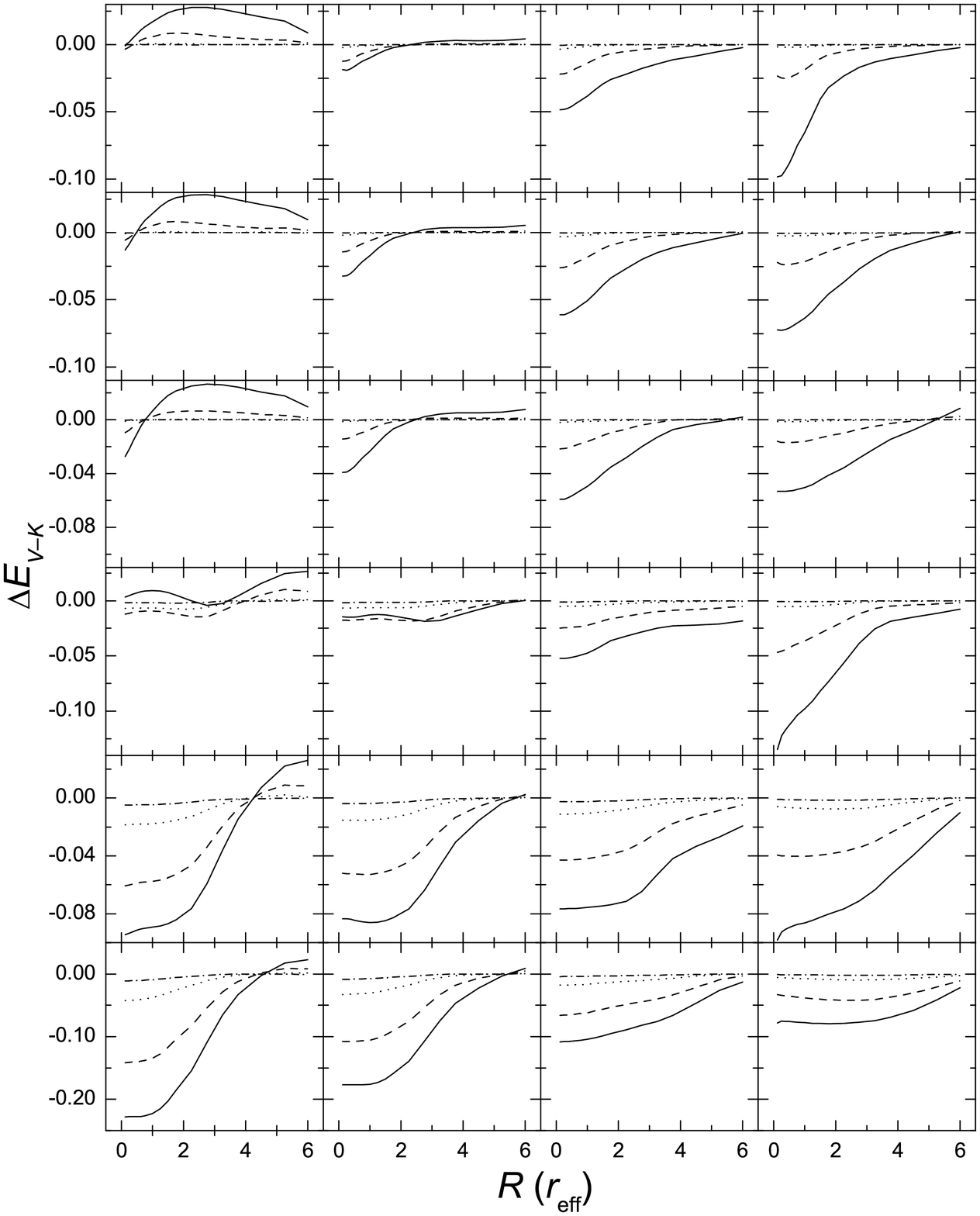,width=124mm,angle=0,clip=}}
\captionb{4}{Differences in
$E_{V-K}$ photometric profile between disk galaxy
models computed with isotropic and anisotropic scattering.
Panel rows correspond to the models S1--S6 (starting from top),
panel columns -- to model inclinations of 0\degr, 50\degr,
80\degr \ and 90\degr \ (left to right). Solid, dashed,
dotted and dash-dotted lines correspond to differences
between the M8:0 model from one side and the M0:8, M1:7,
M3:5 and M5:3 models from other side. The aperture semi-major
axis radius is given in units of the effective stellar disk
scale-length $r_{\rm eff}$.}
\end{figure}

\begin{figure}
\centerline{\psfig{figure=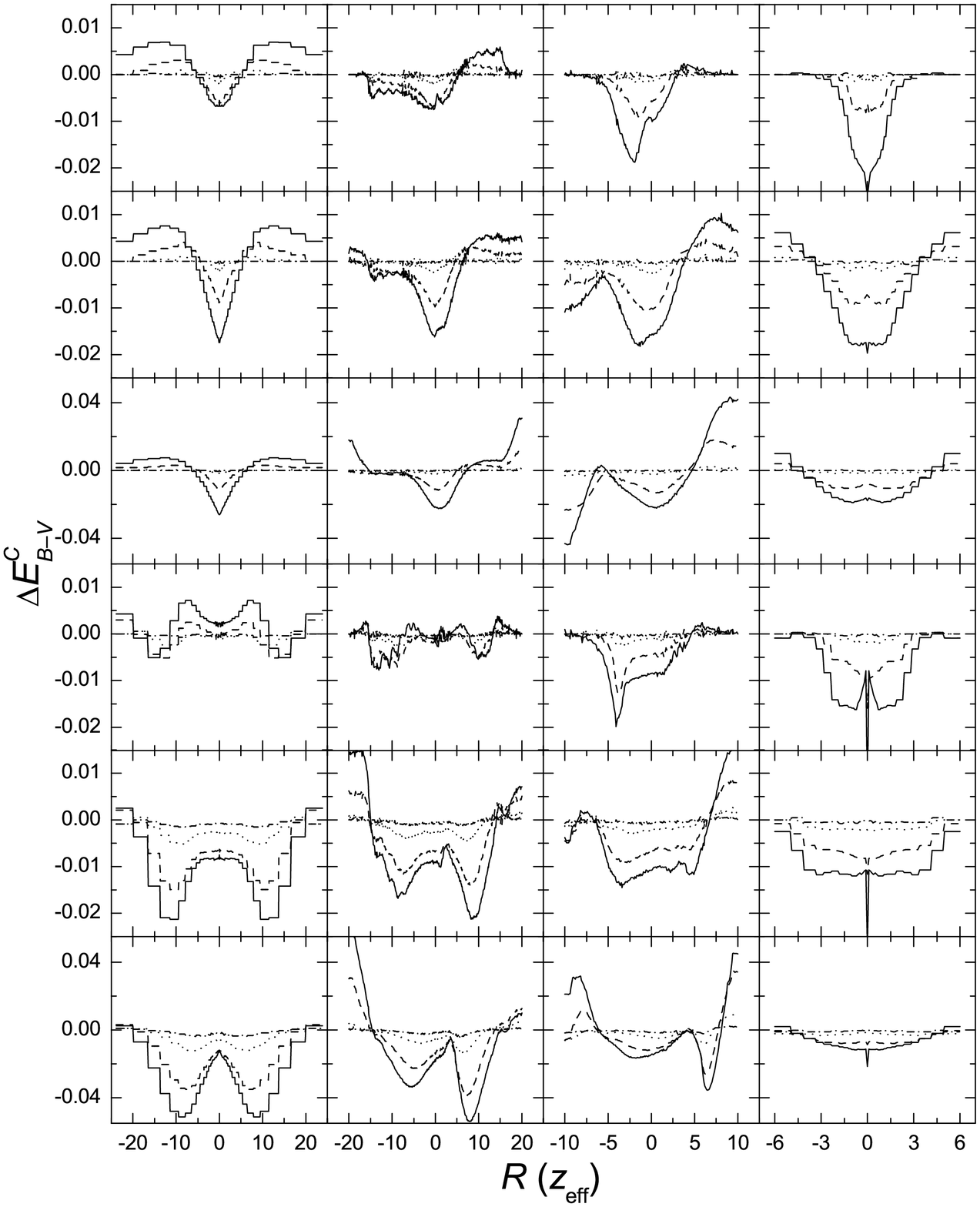,width=124mm,angle=0,clip=}}
\captionb{5}{Differences in
$E_{B-V}$ minor axis cross-section between disk galaxy models
computed with isotropic and anisotropic scattering. Panel rows
correspond to models S1--S6 (starting from top), panel columns -- to
model inclinations of 0\degr, 50\degr, 80\degr\ and 90\degr\ (left to
right). Solid, dashed, dotted and dash-dotted lines correspond
to differences between the M8:0 model from one side and the M0:8,
M1:7, M3:5 and M5:3 models from other side. The distance from the model
galaxy image center is given in units of the effective stellar disk
scale-height $z_{\rm eff}$, negative $R$ values correspond to
half of the model closer to the observer.}
\end{figure}

\begin{figure}
\centerline{\psfig{figure=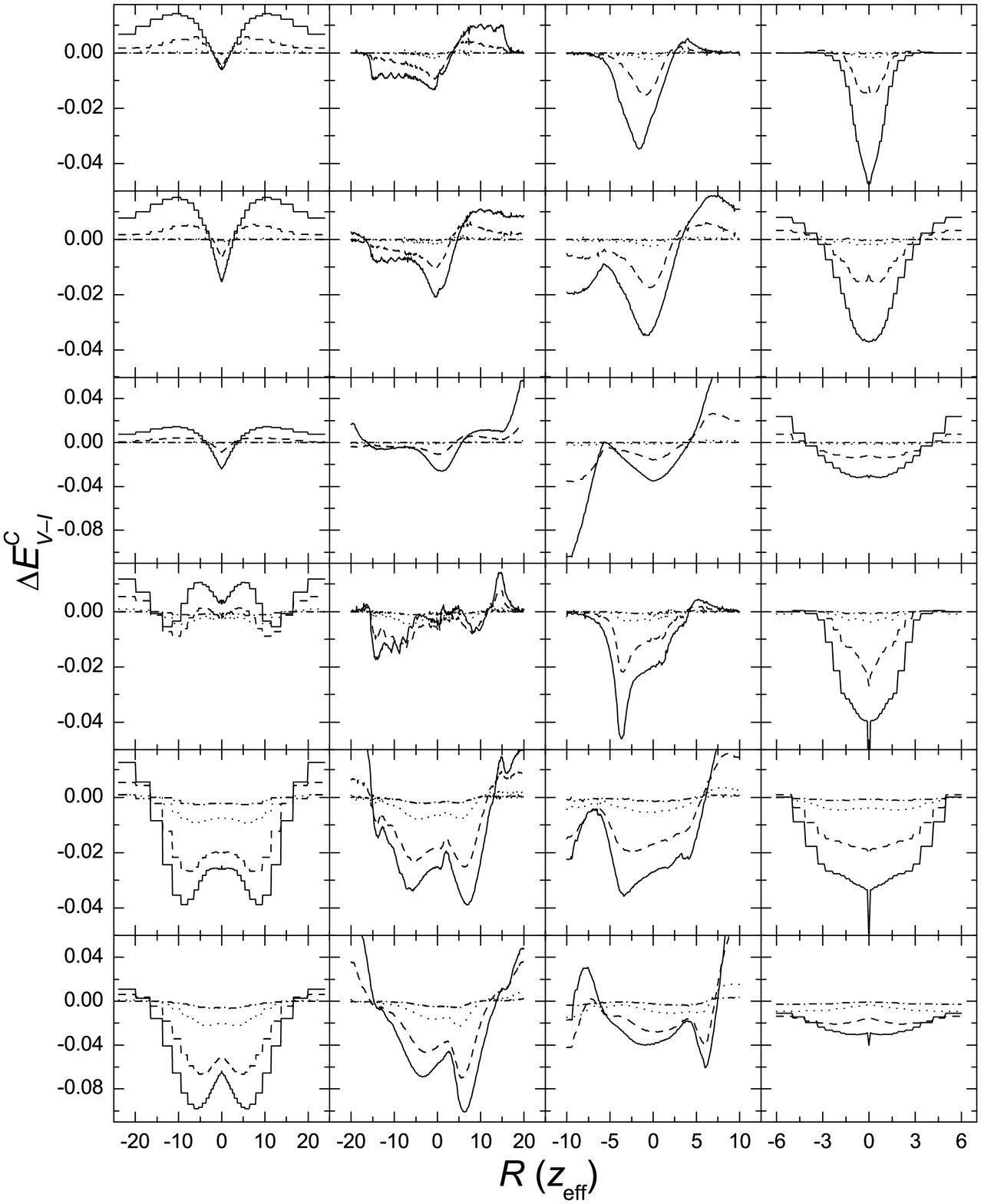,width=124mm,angle=0,clip=}}
\captionc{6}{The same as in Fig. 5, but for
$E_{V-I}$ minor axis cross-section.}
\end{figure}


The effects of anisotropic scattering on the photometric profile of
color excess, as shown in Figure 4 for $E_{V-K}$, are the most prominent
for the models S5 and S6 seen face-on.  The central values of color
excess appear to be overestimated for all models computed with any level
of isotropic scattering approximation.  The model group S1--S4 shows
acceptable accuracy starting from the M1:7 models, approximated
solutions significantly deviate from fully anisotropic cases only at
high inclination, where it would be necessary to apply anisotropic
treatment up to the first three iterations.

\begin{figure}
\centerline{\psfig{figure=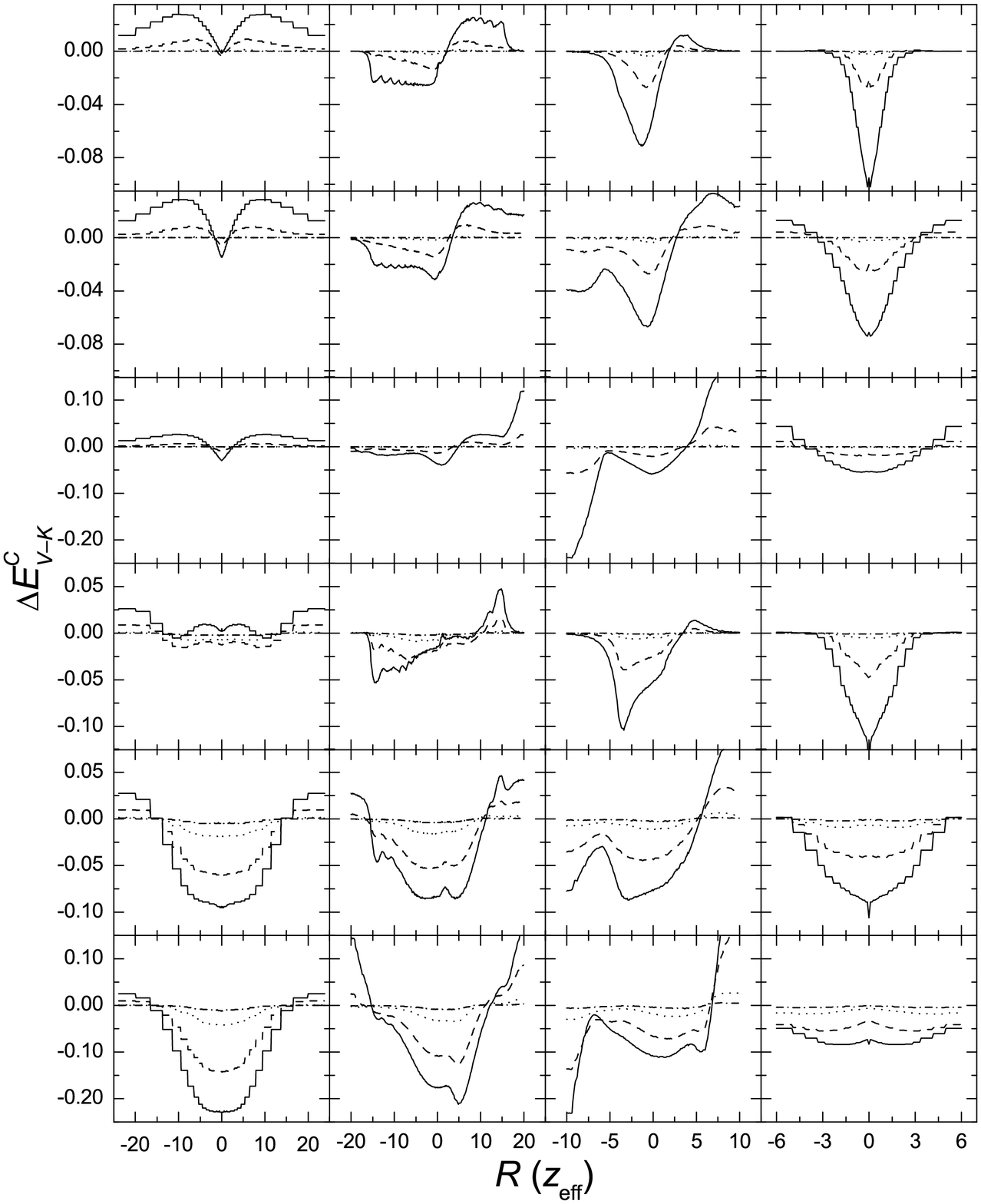,width=124mm,angle=0,clip=}}
\captionc{7}{The same as in Fig. 5, but for
$E_{V-K}$ minor axis cross-section.}
\end{figure}

Differences of color excesses of the minor axis cross-sections, $\Delta
E_{B-V}^C$, $\Delta E_{V-I}^C$ and $\Delta E_{V-K}^C$, computed for the
models with isotropic approximation from one side and for the models
with anisotropic scattering phase function on other side, are shown in
Figures 5--7.  The deviations in $E_{U-V}^C$ and $E_{V-R}^C$ being
identical to $\Delta E_{B-V}^C$ and $\Delta E_{V-I}^C$ in form, differ
from them only by absolute values.  Similarly to the results, obtained
for azimuthally-averaged photometric profiles, two model groups can be
distinguished, low opacity group S1--S4 and high opacity group S5--S6.
Galaxy models in the S1--S4 group computed assuming the isotropic
scattering in most cases display small deviations from the fully
anisotropic M8:0 models, mostly concentrated near the image center.  In
contrast, the model group S5--S6 shows large errors at all inclinations,
affecting significant part of the minor axis cross-sections, thus
requiring at least three (in some cases up to five) first iterations to
fully account for the scattering anisotropy.

As can be seen in Figures 4--7, the purely isotropic models provide a
poor approximation to the solutions obtained with appropriate values of
$g_\lambda$.  The M1:7 approximation offers an improvement by a factor
of 2 or more over M0:8 in reproducing photometric profiles of models in
the S1--S4 group, while for the model families S5 and S6 the improvement
shown by model M1:7 in respect to M0:8 is less prominent.  In general,
the M3:5 models produce the results that with acceptable tolerance can
be considered identical to fully anisotropic models.  Isotropic
approximation in the 6th and subsequent iterations does not introduce a
significant error since these iterations contribute only minor fraction
to the energy balance.  However, when comparing the relative differences
between models M1:7, M3:5 and M5:3, it is necessary to take into account
both the required numerical accuracy and the computational efficiency of
the code.

Approximation of some iterations using the isotropic phase function has
a potential to improve the performance of RT problem-solving code by
reducing computing time and memory requirements.  For Monte-Carlo based
methods such approximation will allow use of more efficient algorithms
as well as decrease resources necessary to follow individual photons
through the model.  For the iterative RT problem-solving algorithms,
such as ray-tracing based code used in this work, the isotropic
approximation also significantly speeds up calculations and improves
code's accuracy, since it eliminates numerical errors caused by
otherwise necessary interpolation of the scattering phase function.

\sectionb{4}{CONCLUSIONS}
\vskip-2mm

We have presented a study of the effect of isotropic scattering phase
function approximation on photometric properties of the model disk
galaxies having several commonly assumed star and dust relative
distributions.  It has been shown that in all cases it is necessary to
fully take into account the anisotropic scattering in at least the first
scattering event.  The subsequent scatterings can be assumed to be
isotropic with varying degree of accuracy depending on the model
extinction distribution.  The effect of anisotropic scattering in terms
of higher than 5th order is negligible since their contributions to the
overal energy balance of the model is small.  The largest errors in the
extinction and color excess profiles and the minor axis cross-sections,
caused by isotropic approximation, were found for the models with dust
distribution being more extended than the stars, while models with the
extinction concentrated at the disk midplane are least affected by the
isotropic scattering approximation.  It was found that the effect of
isotropic approximation of scattering phase function on the color excess
photometric profiles and the minor axis cross-sections depends strongly
on the difference in $g_\lambda$ of the involved photometric passbands,
while the effect of the difference in the relative extinction
$\tau_\lambda / \tau_{\rm V}$ is marginal.

\vskip3mm

ACKNOWLEDGMENT.\ This work has been supported by the Lithuanian
State Science and Studies Foundation.

\References
\vskip-3mm

\baselineskip=10pt
\enlargethispage{4mm}

\refb
Chandrasekhar S. 1960, {\it Radiative Transfer}, New York, Dover
\vskip-0.2mm

\refb
Ferrara A., Bianchi S., Cimatti A., Giovanardi C. 1999, ApJS, 123, 427
\vskip-0.2mm

\refb
Henyey L., Greenstein J. 1941, ApJ, 93, 70
\vskip-0.2mm

\refb
Laor A., Draine B. 1993, ApJ, 402, 441
\vskip-0.2mm

\refb
Li A., Greenberg M. 2003, in {\it Solid State Astrochemistry},
eds. V. Pirronello, J. Krelowski \& G. Manic{\`o},
Kluwer Academic Publishers, Dordrecht, p. 37
\vskip-0.2mm

\refb
Semionov D., Vansevi{\v c}ius V. 2002, Baltic Astronomy, 11, 537
\vskip-0.2mm

\refb Semionov D. 2003, {\it Spectrophotometric Evolution of Dusty Disk
Galaxies}, Inst. of Theoretical Physics and Astronomy, Vilnius, PhD
thesis
\vskip-0.2mm

\refb
Semionov D., Vansevi{\v c}ius V. 2005, astro-ph/0501146
\vskip-0.2mm

\refb
Silva L., Granato G., Bressan A., Danese L. 1998, ApJ, 509, 103

\end{document}